\documentclass[pre,preprint,english,showpacs,aps,prb,a4paper,floatfix]{revtex4}
\usepackage[T1]{fontenc}
\usepackage{color}
\usepackage[latin1]{inputenc}
\usepackage{graphicx}
\usepackage{bm}
\usepackage{epsfig}

\begin{document}

\title{Phase behavior of liquid-crystal monolayers of rod-like and plate-like particles}

\author{Yuri Mart\'{\i}nez-Rat\'on}
\affiliation{Grupo Interdisciplinar de Sistemas Complejos (GISC), Departamento de Matem\'aticas, Escuela Polit\'ecnica Superior,
Universidad Carlos III de Madrid, Avenida de la Universidad 30, E-28911, Legan\'es, Madrid, Spain}
\email{yuri@math.uc3m.es}

\author{Szabolcs Varga}
\affiliation{Institute of Physics and Mechatronics, University of Pannonia, PO Box 158, Veszpr\'em, H-8201 Hungary}
\email{vargasz@almos.uni-pannon.hu}

\author{Enrique Velasco}
\affiliation{Departamento de F\'{\i}sica Te\'orica de la Materia Condensada, Instituto de Ciencia de Materiales Nicol\'as 
Cabrera and Condensed Matter Physics Center (IFIMAC), Universidad Aut\'onoma de Madrid, E-28049 Madrid, Spain}
\email{enrique.velasco@uam.es}

\begin{abstract}
Orientational and positional ordering properties of liquid crystal monolayers are
examined by means of Fundamental-Measure Density Functional Theory. Particles forming the monolayer are
modeled as hard parallelepipeds of square section of size $\sigma$ and length $L$. Their shapes are controlled by 
the aspect ratio $\kappa=L/\sigma$ ($>1$ for prolate and $<1$ for oblate shapes).  
The particle centers of mass are restricted to a flat surface and three possible and mutually perpendicular orientations
(in-plane and along the layer normal) of their uniaxial axes are allowed. We find that the structure of the monolayer
depends strongly on particle shape and density. In the case of rod-like shapes, particles align along the layer normal 
in order to achieve the lowest possible occupied area per particle. This phase is a uniaxial nematic
even at very low densities. In contrast, for plate-like particles, the lowest occupied area can be achieved
by random in-plane ordering in the monolayer, i.e. planar
nematic ordering takes place even at vanishing densities. It is found that the random
in-plane ordering is not favorable at higher densities and the system undergoes an
in-plane ordering transition forming a biaxial nematic phase or crystallizes. For certain
values of the aspect ratio, the uniaxial-biaxial nematic phase transition is observed for
both rod-like and plate-like shapes. The stability region of the biaxial nematic phase
enhances with decreasing aspect ratios for plate-like particles, while the rod-like
particles exhibit a reentrant phenomenon, i.e. a sequence of uniaxial-biaxial-uniaxial nematic ordering
with increasing density if the aspect ratio is larger than 21.34. In addition to
this, packing fraction inversion is observed with increasing surface pressure due to the 
alignment along the layers normal. At very high densities the nematic phase destabilizes
to a nonuniform phases (columnar, smectic or crystalline phases) for both shapes. 
\end{abstract}


\maketitle

\section{Introduction}
Understanding the ordering properties of liquid crystals in restricted geometry is still a
challenging problem \cite{1}. It is well known that the confinement of rod-like or plate-like
particles into two dimensions has great impact on the phase behaviour of the system. For
example, the isotropic-nematic phase transition of three-dimensional (3D) hard ellipsoids
is of first order \cite{2}, but when the centers of mass and orientations of the
long axis of the particles are restricted to be on a plane, which corresponds to a two-dimensional 
(2D) system of hard ellipses, a continuous isotropic-nematic phase transition can be observed (via a
Kosterlitz-Thouless disclination unbinding type mechanism with a nematic phase
exhibiting only quasi-long-range orientational order \cite{3}). The defect structure of a 2D
nematic phase is also nontrivial \cite{4,5,6}. In addition to this, the dynamics of 2D hard
ellipses shows peculiarities in the rotational and translational diffusions and in the
glassy behaviours \cite{7,8}. Recently, a two-step glass transition has been observed in a
monolayer of colloidal ellipsoids confined between two glass walls. In the first step, an
orientational glass emerges with increasing density, which is related to orientational
arrest of the particles, while in the second the system becomes translationally
frozen at higher densities \cite{9,10}.
The structural and dynamical properties of 2D and even 1D complex fluids have recently become
experimentally accessible due to the development of nanofluidics and optical
trapping methods. Liquid-crystal monolayers can be prepared by confining colloidal
particles between parallel walls \cite{11,12} or by spreading colloidal nanoparticles or
amphiphilic molecules at the air/liquid interface \cite{13,14}. This makes it possible to study
the competition between the orientational and packing entropies in restricted geometries
\cite{12}. Even the ordering effect of surface patchiness has been examined in a monolayer
of nanoplatelets \cite{15}.

The difference between the phase behaviours of molecularly thick films (Langmuir
monolayers) and that of colloidal monolayers confined between two parallel planes is
mainly due to the fact that the amphiphilic molecules are allowed to rotate out from the
air/liquid interface, while the colloids can rotate only in the plane parallel to the confining walls.
The consequence is that the phase behaviour of Langmuir monolayers can be much richer
than that of 2D colloidal systems. For example, upon compression, only a few phases (isotropic, nematic,
solid) may occur if the confined particles have only 2D
orientational freedom \cite{6,17}, while several additional, tilted or not tilted,
phases can be present in the case of Langmuir monolayers with out-of-plane rotational freedom \cite{18,19,20}
due to their intermediate quasi-two-dimensional character (i.e. 2D in translations and 3D in orientations). 
Therefore it is worth studying the effect of the orientational freedom and its coupling with the translational
part on the ordering properties of simple model systems. A model of confined hard particles is ideal, since it
is amenable to theoretical analysis and, as is well known from numerous studies on liquid-crystalline ordering,
can give rise to nontrivial behaviour. 

In our study we examine the phase behaviour of uniaxial hard parallelepipeds confined such that their
centers of mass are forced to be on a plane, while they are allowed to rotate out of the plane. 
Our aim is to determine the effect of the additional, out-of-plane, orientational freedom on the stability of the 
isotropic, nematic and solid phases of 2D hard rectangles. 
Depending on the aspect ratio of the
parallelepipeds, it is possible to study the phase behaviour of both rod-like (prolate
shaped) and plate-like (oblate shaped) particles. In the case of rod-like particles,
steric (excluded volume) interactions favour orientational ordering along the layer
normal even at low densities, because out-of-plane ordering produces low surface
coverage. However, the situation is very different in the case of oblate shapes, because when particles
lie on the plane the occupied area on the surface is increased. Therefore, steric interactions act on the plate-like particles so as to promote
rotation out of the confining plane. In such situation biaxial nematic ordering may emerge in
the monolayer, because both symmetry axes of the particles can be ordered at high surface
coverage. We pay special attention to the determination of the stability region of
biaxial nematic phase for both shapes. Note that the biaxial nematic phase has been only
stabilized in systems of biaxial hard particles \cite{21,22,23,24}, while our model system is
uniaxial and the confinement gives rise to biaxial ordering.

The article is arranged as follows: Sec. \ref{Zwanzig} is devoted to presenting the model for the monolayer and 
the density-functional theory (DFT) and related tools used to calculate phase diagrams; in Sec. \ref{nematic} we focus on 
the nematic phases, while in Sec. \ref{bifurc} we present details on the bifurcation analysis used to calculate the spinodal 
instabilities of the nematic phase with respect to nonuniform phases. The results are shown in Sec. 
\ref{results} for prolate (Sec. \ref{prolate}) and oblate (Sec. \ref{oblate}) particle geometries. Some conclusions are 
drawn in Sec. \ref{conclusions}. 

\section{Zwanzig model for 3D hard rods projected on a plane}
\label{Zwanzig}

\begin{figure}
\epsfig{file=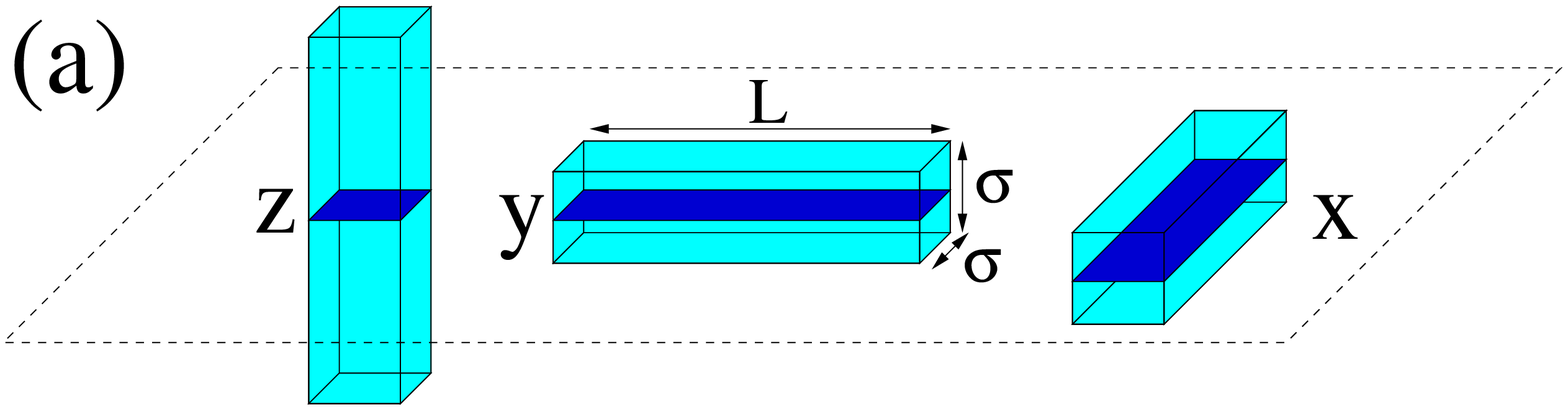,width=3.in}
\epsfig{file=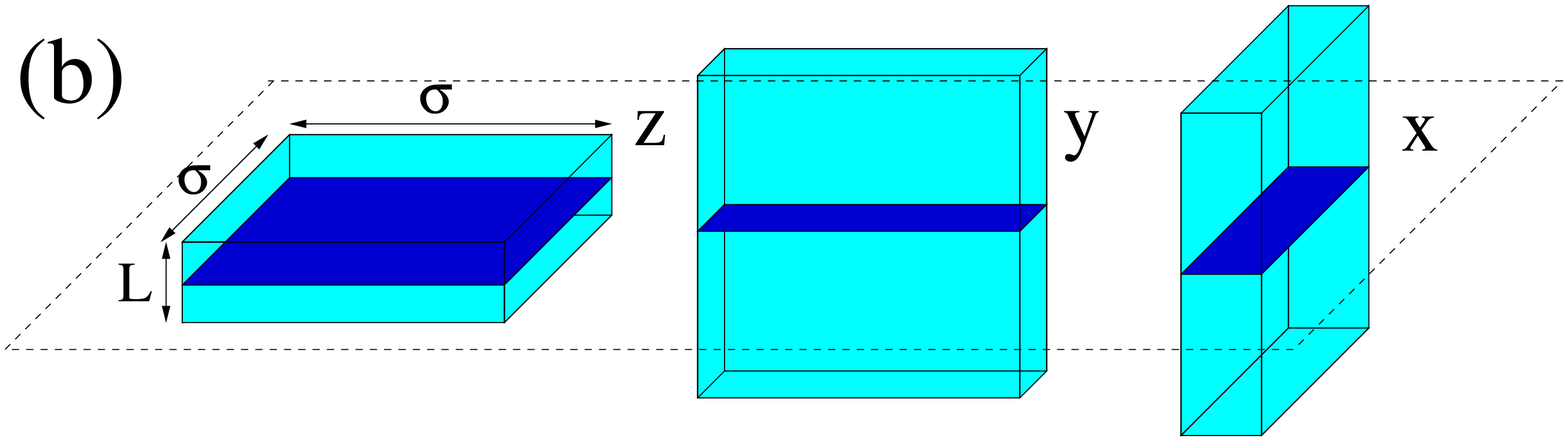,width=3.in}
\caption{Possible orientations of confined (a) prolate and (b) oblate uniaxial parallelepipeds in the Zwanzig 
approximation. The projection of the original particles onto a plane are shown in dark. Different species are 
correspondingly labelled and their characteristic lengths are also shown.}
\label{fig1}
\end{figure}

We use a Zwanzig approximation, where particles are restricted to orient along $x$, $y$ and $z$ axes only.
The model is defined by three parallelepipedic species of length $L$ and square cross-sectional area 
$\sigma^2$ ($\sigma$ being the particle breadth). The aspect ratio $\kappa=L/\sigma$ can be larger (prolate geometry) or 
smaller (oblate geometry) than unity. The three species are labelled as $x$, $y$ and $z$, meaning that the 
longest (prolate shape) and shortest (oblate shape) axis of the particles is parallel to the $x$, $y$ or $z$ Cartesian axes. 
The particle centers of mass are restricted to lie
on a plane perpendicular to $z$ and located at $z=0$. Therefore, the 3D density profiles are 
\begin{eqnarray}
\rho_{\nu}^{(\rm 3D)}(x,y,z)=\rho^{(\rm 2D)}_{\nu}(x,y)\delta(z),\quad  \nu=x,y,z,
\label{cero}
\end{eqnarray}
with $\delta(z)$ the Dirac-delta function. $\rho^{(\rm 2D)}_{\nu}(x,y)$ is the 2D density profile of species $\nu$ 
on the plane. There are three possible projections of the particles on the plane (Fig. \ref{fig1}): two rectangles, 
with their distinct axis parallel to $x$ or $y$, of width $\sigma$ and length $L$, and a square of side-length $\sigma$.
Thus our model becomes effectively a 2D ternary mixture of two species of mutually perpendicular hard 
rectangles and a third species consisting of hard squares. 

The DFT theory for the 2D ternary mixture model defined above can be obtained from the Fundamental-Measure DFT expression
for a system of 3D hard parallelepipeds in the Zwanzig approximation, as we now show. The excess part of the latter has 
the form
\begin{eqnarray}
{\cal F}^{(\rm 3D)}_{\rm ex}[\{\rho_{\nu}^{(\rm 3D)}\}]=\int dx \int dy \int dz \ \Phi_{\rm 3D}(x,y,z),
\label{ala}
\end{eqnarray}
where $\Phi_{\rm 3D}(x,y,z)$ is the excess part of the free-energy density (see \cite{Cuesta1} for details). This functional 
fulfills the dimensional crossover property: when the constrained density profiles (\ref{cero}) are inserted into (\ref{ala}), 
one obtains the free-energy functional of the 2D projected fluid:  
\begin{eqnarray}
{\cal F}^{(\rm 3D)}_{\rm ex}[\{\rho_{\nu}^{(\rm 3D)}\}] \to 
{\cal F}^{(\rm 2D)}_{\rm ex}[\{\rho_{\nu}^{(\rm 2D)}\}]=\int dx \int dy \ \Phi_{\rm 2D}(x,y),
\label{2D}
\end{eqnarray}
with $ \Phi_{\rm 2D}(x,y)$ the excess part of the free-energy density of the 2D ternary mixture \cite{Cuesta1}:
\begin{eqnarray}
\Phi_{\rm 2D}({\bm r})=-n_0(\bm r)\ln[1-n_2(\bm r)]+\frac{n_{1x}(\bm r)n_{1y}(\bm r)}{1-n_2(\bm r)},
\end{eqnarray}
where the notation ${\bm r}\equiv (x,y)$ has been used. The weighted densities are given by 
\begin{eqnarray}
n_{\alpha}(\bm r)=\sum_{\nu=\{x,y,z\}}\int d\bm r' \rho_{\nu}^{(\rm 2D)}(\bm r')\omega^{(\alpha)}_{\nu}(\bm r-\bm r')
\end{eqnarray}
with
\begin{eqnarray}
\omega^{(0)}_{\nu}(\bm r)&=&\frac{1}{4}\delta\left(\frac{\sigma^{x}_{\nu}}{2}-|x|\right)
\delta\left(\frac{\sigma^{y}_{\nu}}{2}-|y|\right),\nonumber\\\nonumber\\
\omega^{(1x)}_{\nu}(\bm r)&=&\frac{1}{2}\Theta\left(\frac{\sigma^{x}_{\nu}}{2}-|x|\right)
\delta\left(\frac{\sigma^{y}_{\nu}}{2}-|y|\right),\nonumber\\\nonumber\\
\omega^{(1y)}_{\nu}(\bm r)&=&\frac{1}{2}\delta\left(\frac{\sigma^{x}_{\nu}}{2}-|x|\right)
\Theta\left(\frac{\sigma^{y}_{\nu}}{2}-|y|\right),\nonumber\\\nonumber\\
\omega^{(2)}_{\nu}(\bm r)&=&\Theta\left(\frac{\sigma^{x}_{\nu}}{2}-|x|\right)\Theta\left(\frac{\sigma^{y}_{\nu}}{2}-|y|\right),
\end{eqnarray}
$\Theta(x)$ is the Heaviside function and we have defined $\sigma^{\mu}_{\nu}\equiv\sigma+(L-\sigma)\delta_{\mu\nu}$, 
with $\delta_{\mu\nu}$ the Kronecker delta. Note that $\mu=\{x,y\}$ while $\nu=\{x,y,z\}$. The ideal part of the free-energy 
is
\begin{eqnarray}
\beta {\cal F}_{\rm id}[\{\rho_{\mu}\}]=\sum_{\nu=\{x,y,z\}} \int d{\bm r} 
\rho_{\nu}^{(2D)}(\bm r)\left[\ln \rho_{\nu}^{(2D)}(\bm r) -1\right].
\label{ideal}
\end{eqnarray}
Note that we have dropped the thermal volume $\Lambda^3$ inside the logarithm of Eq. (\ref{ideal}) 
as it does not affect the phase behaviour of the system.

\subsection{The nematic phase}
\label{nematic}

For uniform density profiles we have $\rho_{\nu}^{(\rm 2D)}=\gamma_{\nu}\rho_{\rm 2D}$, with $\rho_{\rm 2D}=N/A$ the total 
surface density and $\gamma_{\nu}$ the molar fraction of species $\nu$, which satisfies the constraint
$\displaystyle\sum_{\nu}\gamma_{\nu}=1$. For a general biaxial nematic phase, N$_{\rm b}$, we parameterize $\gamma_{\nu}$ as
\begin{eqnarray}
\gamma_x=\frac{1-Q_{\rm u}}{3}+\frac{Q_{\rm b}}{2},\quad
\gamma_y=\frac{1-Q_{\rm u}}{3}-\frac{Q_{\rm b}}{2},\quad
\gamma_z=\frac{1+2Q_{\rm u}}{3},
\end{eqnarray}
where $Q_{\rm u}\in[-1/2,1]$ is the usual uniaxial nematic order parameter and $Q_{\rm b}\equiv \gamma_x-\gamma_y$ measures 
the biaxiality in the $x-y$ plane. The uniaxial nematic phase N$_u$ has $Q_{\rm b}=0$.
In terms of the order parameters, the ideal and excess parts of the free-energy density 
in reduced units become
\begin{eqnarray}
&&\frac{\beta{\cal F}_{\rm id}a}{V}=\rho^*\left[\ln \rho^*-1+\sum_{\nu=x,y,z}\gamma_{\nu}\left(Q_{\rm u},Q_{\rm b}\right)
\ln \gamma_{\nu}\left(Q_{\rm u},Q_{\rm b}\right)\right],\nonumber\\
&&\frac{\beta{\cal F}_{\rm ex}a}{V}=-\rho^*\ln\left[1-\eta(Q_{\rm u})\right]\nonumber\\&&\hspace{4cm}
+\frac{\kappa\left(\rho^*\right)^2}{1-\eta(Q_{\rm u})}
\left[\frac{1}{9}\left(1+\frac{2}{\kappa}-\left(1-\frac{1}{\kappa}\right)Q_{\rm u}\right)^2-\frac{1}{4}
\left(1-\frac{1}{\kappa}\right)^2Q_{\rm b}^2\right],
\end{eqnarray}
with $V$ the total volume of the system, and $\rho^*=\rho_{\rm 2D}a$ (with $a=L\sigma$ the area of the rectangular 
species $x$ and $y$) a scaled 2D density and $\kappa=L/\sigma$ the aspect ratio. The uniform limit of the weighted
density $n_2(\bm r)$ is the packing fraction $\eta=\displaystyle\frac{N}{A}\sum_{\nu}\gamma_{\nu}a_{\nu}$ 
(with $a_x=a_y=L\sigma$ and $a_z=\sigma^2$), 
which depends on $Q_{\rm u}$:
\begin{eqnarray}
\eta(Q_{\rm u})=\frac{\rho^*}{3}\left[2+\frac{1}{\kappa}-2\left(1-\frac{1}{\kappa}\right)Q_{\rm u}\right].
\label{the_packing}
\end{eqnarray}
Minimization of the total free energy $\beta{\cal F}a/V=\beta\left({\cal F}_{\rm id}+{\cal F}_{\rm ex}\right)a/V$ with 
respect to the order parameters  $Q_{\rm u}$ and $Q_{\rm b}$ gives a pair of nonlinear equations that have to be
solved iteratively:
\begin{eqnarray}
Q_{\rm u}&=&\frac{1-e^{-\displaystyle{H_1(Q_{\rm u},Q_{\rm b})}}\cosh\left[H_2(Q_{\rm u},Q_{\rm b})\right]}{
1+2e^{-\displaystyle{H_1(Q_{\rm u},Q_{\rm b})}}\cosh\left[H_2(Q_{\rm u},Q_{\rm b})\right]},
\nonumber\\\nonumber\\
Q_{\rm b}&=&\frac{2e^{-\displaystyle{H_1(Q_{\rm u},Q_{\rm b})}}\sinh\left[H_2(Q_{\rm u},Q_{\rm b})\right]}
{1+2e^{-\displaystyle{H_1(Q_{\rm u},Q_{\rm b})}}\cosh\left[H_2(Q_{\rm u},Q_{\rm b})\right]}\label{dos},
\end{eqnarray}
where we have defined
\begin{eqnarray}
&&H_1(Q_{\rm u},Q_{\rm b})=\frac{\rho^*}{1-\eta(Q_{\rm u})}\left(1-\frac{1}{\kappa}\right)\left\{
1+\frac{\kappa}{3}\left(1+\frac{2}{\kappa}-\left(1-\frac{1}{\kappa}\right)Q_{\rm u}\right)\right.\nonumber\\\nonumber\\
&&\hspace{2.2cm}\left. +\frac{\kappa\rho^*}{1-\eta(Q_{\rm u})}\left[
\frac{1}{9}\left(1+\frac{2}{\kappa}-\left(1-\frac{1}{\kappa}\right)Q_{\rm u}\right)^2-\frac{1}{4}
\left(1-\frac{1}{\kappa}\right)^2Q_{\rm b}^2\right]\right\},\nonumber\\\nonumber\\
&&H_2(Q_{\rm u},Q_{\rm b})=\frac{\kappa \rho^*}{2[1-\eta(Q_{\rm u})]}\left(1-\frac{1}{\kappa}\right)^2Q_{\rm b}.
\end{eqnarray}
The N$_u$-N$_b$ bifurcation can be calculated by solving Eqns. (\ref{dos}), expanded up to first order 
with respect to $Q_{\rm b}$. After some algebra, we arrive at the non-linear equation 
\begin{eqnarray}
f(y,\kappa)\equiv y(\kappa^2-1)-2-\exp{\left\{\kappa\left[\frac{\left(y(1+\kappa^{-1})+1\right)^2}{1-\kappa^{-1}}-1\right]\right\}}=0.
\label{bifurcation}
\end{eqnarray}
for the variable $y\equiv \rho_{\rm 2D}\sigma^2/(1-\rho_{\rm 2D}\sigma^2)$ [note that $\rho^*=y\kappa/(1+y)$].
One obtains a continuous N$_{\rm u}$-N$_{\rm b}$ phase transition, with a phase boundary in the density--aspect ratio
plane given by a curve $y(\kappa)$ or $\rho^*(\kappa)$; the uniaxial order parameter $Q_{\rm u}$ and the packing fraction 
$\eta$ at bifurcation can be calculated as a function of $y$ as
\begin{eqnarray}
Q_{\rm u}=1-\frac{3}{y(\kappa^2-1)},\label{ordering}\hspace{0.6cm}
\eta=(1+y)^{-1}\left(y+\frac{2}{\kappa+1}\right). \label{packk}
\end{eqnarray}
The asymptotic limit $\kappa\to 0$ (infinitely thin plates) of Eq. (\ref{bifurcation}) is $|y|-2=e^{-y^2}$, the
solution $y_a$ of which gives  
$\eta_{\rm a}= (|y_a|-2)/(|y_a|-1)=0.01681$.  
The other asymptotic limit of (\ref{bifurcation}) for prolate particles, $\kappa\to\infty$ (needles), will be discussed 
later.

The critical end-point $(y_0,\kappa_0)$ of the transition curve N$_u$-N$_b$, if it exists, can be found from the 
solution of the equations
\begin{eqnarray}
f(y_0,\kappa_0)=\frac{\partial f}{\partial y}(y_0,\kappa_0)=0,
\end{eqnarray}
which results in the following trascendental equation:
\begin{eqnarray}
\frac{1}{4y_0(1+y_0)^2}=\exp{\left\{\frac{(1+2y_0)(3+2y_0)[2+(3+2y_0)y_0]}{2(1+y_0)}-1\right\}},
\quad \kappa_0=1+\frac{1+2y_0}{2y_0(1+y_0)}.\nonumber\\
\label{tres}
\end{eqnarray}
Solving for prolate parallelepipeds ($\kappa>1$), we obtain the solutions $\rho^*_0=0.52427$ and $\kappa_0=21.33910$.
No solution has been found for oblate parallelepipeds.

\subsection{The nematic to non-uniform phases instabilities}
\label{bifurc}

The Fourier transforms of the direct correlation functions can be written as
\begin{eqnarray}
-\hat{c}_{\mu\nu}({\bm q},\rho^*,\{Q_{\alpha}\})=
\sum_{\alpha,\beta}\frac{\partial^2\Phi_{\rm 2D}(\rho^*,\{Q_{\alpha}\})}{\partial n_{\alpha}\partial n_{\beta}}
\hat{\omega}_{\mu}^{(\alpha)}(\bm q)\hat{\omega}_{\nu}^{(\beta)}(\bm q),
\end{eqnarray}
where $\bm q=(q_x,q_y)$ is the wave vector. The Fourier transforms of the weighting functions are given by
\begin{eqnarray}
\hat{\omega}^{(0)}_{\nu}(\bm q)&=&\chi_0\left(q_x\frac{\sigma_{\nu}^x}{2}\right)\chi_0\left(q_y\frac{\sigma_{\nu}^x}{2}\right),\hspace{0.4cm}
\hat{\omega}^{(1x)}_{\nu}(\bm q)=\sigma_{\nu}^x\chi_1\left(q_x\frac{\sigma_{\nu}^x}{2}\right)\chi_0\left(q_y\frac{\sigma_{\nu}^y}{2}\right),\nonumber\\\nonumber\\
\hat{\omega}^{(1y)}_{\nu}(\bm q)&=&\sigma_{\nu}^y\chi_0\left(q_x\frac{\sigma_{\nu}^x}{2}\right)\chi_1\left(q_y\frac{\sigma_{\nu}^y}{2}\right),\hspace{0.4cm}
\hat{\omega}^{(2)}_{\nu}(\bm q)=\sigma_{\nu}^x\sigma_{\nu}^y\chi_1\left(q_x\frac{\sigma_{\nu}^x}{2}\right)\chi_1\left(q_y\frac{\sigma_{\nu}^y}{2}\right),
\end{eqnarray}
with $\chi_0(x)=\cos x$ and $\chi_1(x)=\sin(x)/x$. The instabilities of the nematic phase against spatially-nonuniform
fluctuations can be found from the equations
\begin{eqnarray}
T({\bm q},\rho^*,\{Q_{\alpha}\})=0,\hspace{0.4cm}\boldsymbol{\nabla}_{\bm q} T({\bm q},\rho^*,\{Q_{\alpha}\})=0,
\label{structure}
\end{eqnarray}
where 
\begin{eqnarray}
T({\bm q},\rho^*,\{Q_{\alpha}\})=\text{det}\left[\frac{\delta_{\mu\nu}}{\rho_{\mu}}-\hat{c}_{\mu\nu}(\bm q,\rho^*,\{Q_{\alpha}\})\right], 
\quad \mu,\nu=x,y,z,
\end{eqnarray}
and `det' denotes a determinant. 
Thus we obtain the values of $\rho^*_0$ and $\bm q_0$ at 
bifurcation, and the order parameters $\{Q_{\alpha}\}$ come from Eqns. (\ref{dos}). 
In practical terms we proceed by fixing the direction of the wavevector as follows: (i) ${\bm q}=(0,q)$, which
implies columnar (C) symmetry if $Q_{\rm b}>0$ or plastic (K) symmetry if $Q_{\rm b}=0$; (ii) ${\bm q}=(q,0)$,
which implies smectic (S) symmetry if $Q_{\rm b}> 0$. The C phase is defined as a 2D layered phase in which the 2D nematic director is 
parallel to the layers (or columns) while in the S phase the director is perpendicular to the layers.  
The value of $q$ can be obtained by solving Eqns. (\ref{structure}).

\section{Results}
\label{results}

In this section we present the results for the model as obtained from (i) minimization of the  
free-energy density and (ii) bifurcation analysis of the continuous  
N$_{\rm u}$-N$_{\rm b}$ transitions for rods and plates or the spinodal instabilities from uniform 
to nonuniform phases. The phase behaviour of prolate and oblate particles are explained in different sections.

\subsection{Prolate particles}
\label{prolate}

\begin{figure}
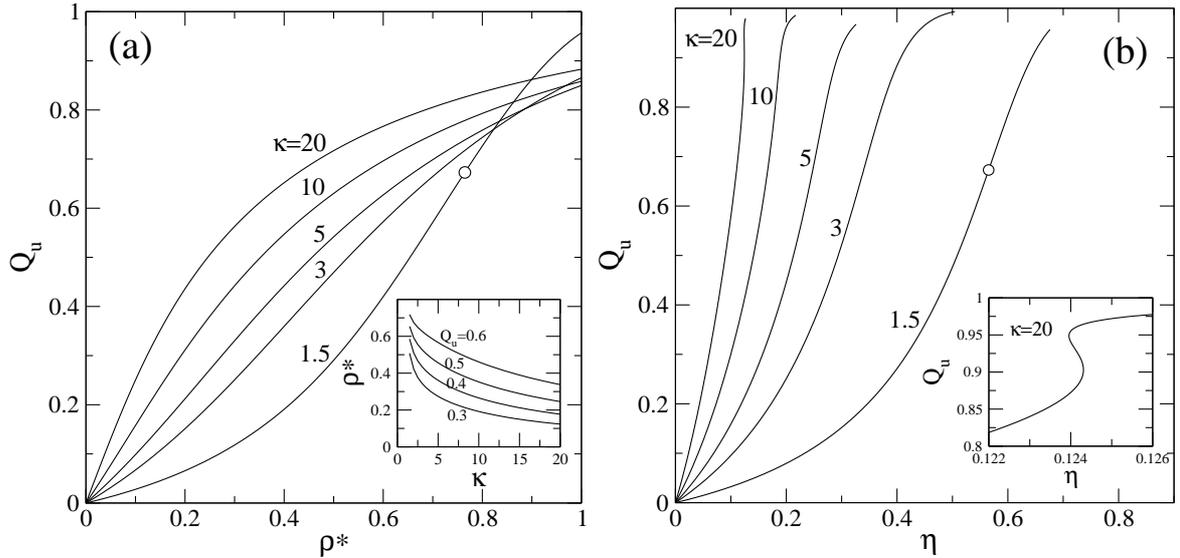

\epsfig{file=fig2a.eps,width=3.in}
\epsfig{file=fig2b.eps,width=3.in}
\caption{(a) Uniaxial parameter $Q_{\rm u}$ as a function of the scaled density $\rho^*$ for different values of $\kappa$ 
(labelled in the figure). Inset: $\rho^*$ as a function of $\kappa$ for various values of $Q_{\rm u}$ (correspondingly labelled). 
(b) $Q_{\rm u}$ as a function of packing fraction $\eta$. Inset: detail of the curve corresponding 
to $\kappa=20$. The empty circle in both panels represents the location of the N$_{\rm u}$-C,K 
bifurcation as calculated from the bifurcation analysis.}
\label{fig2}
\end{figure}

\begin{figure}
\epsfig{file=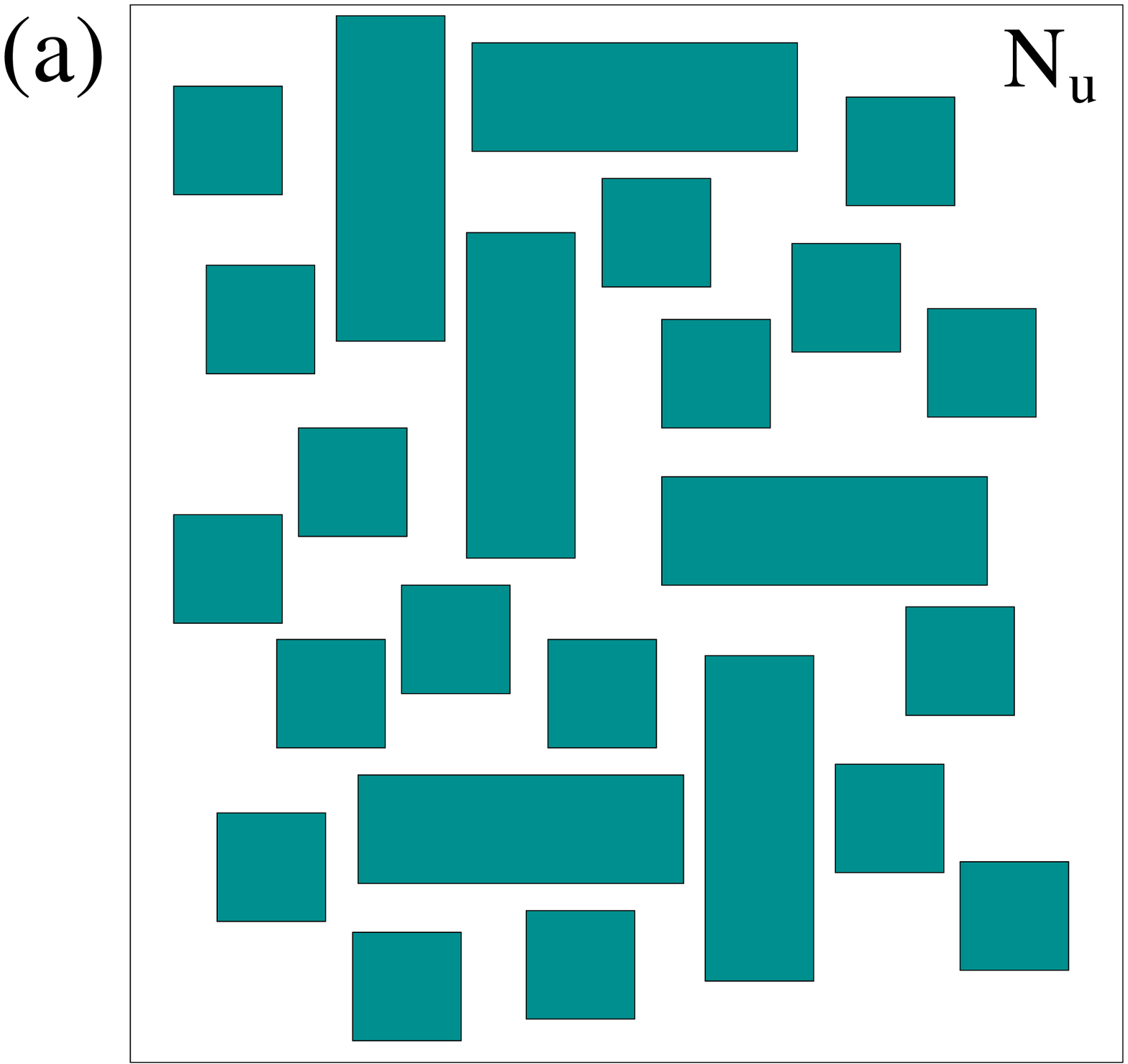,width=1.5in}
\hspace*{0.5cm}
\epsfig{file=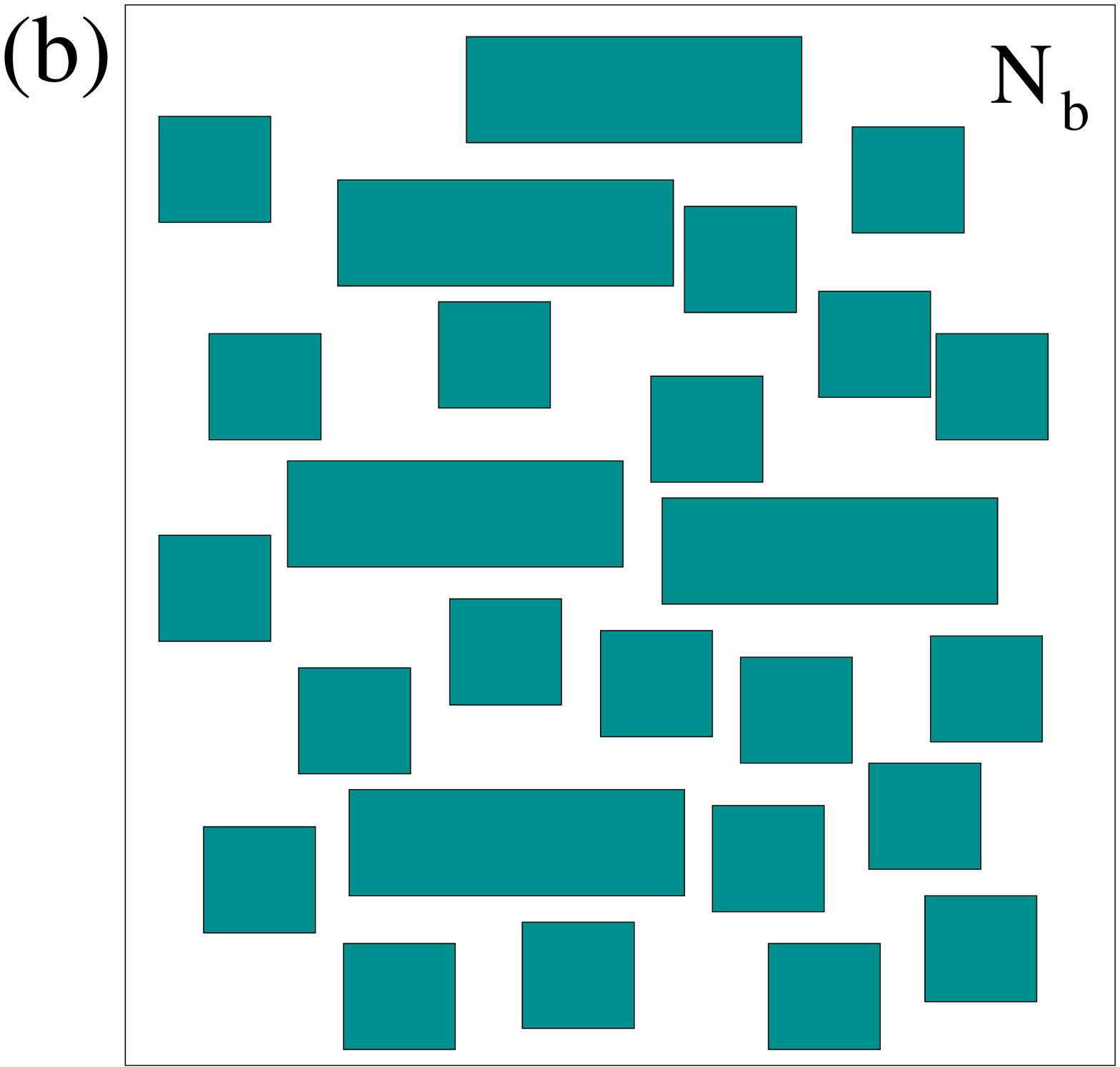,width=1.5in}\\

\epsfig{file=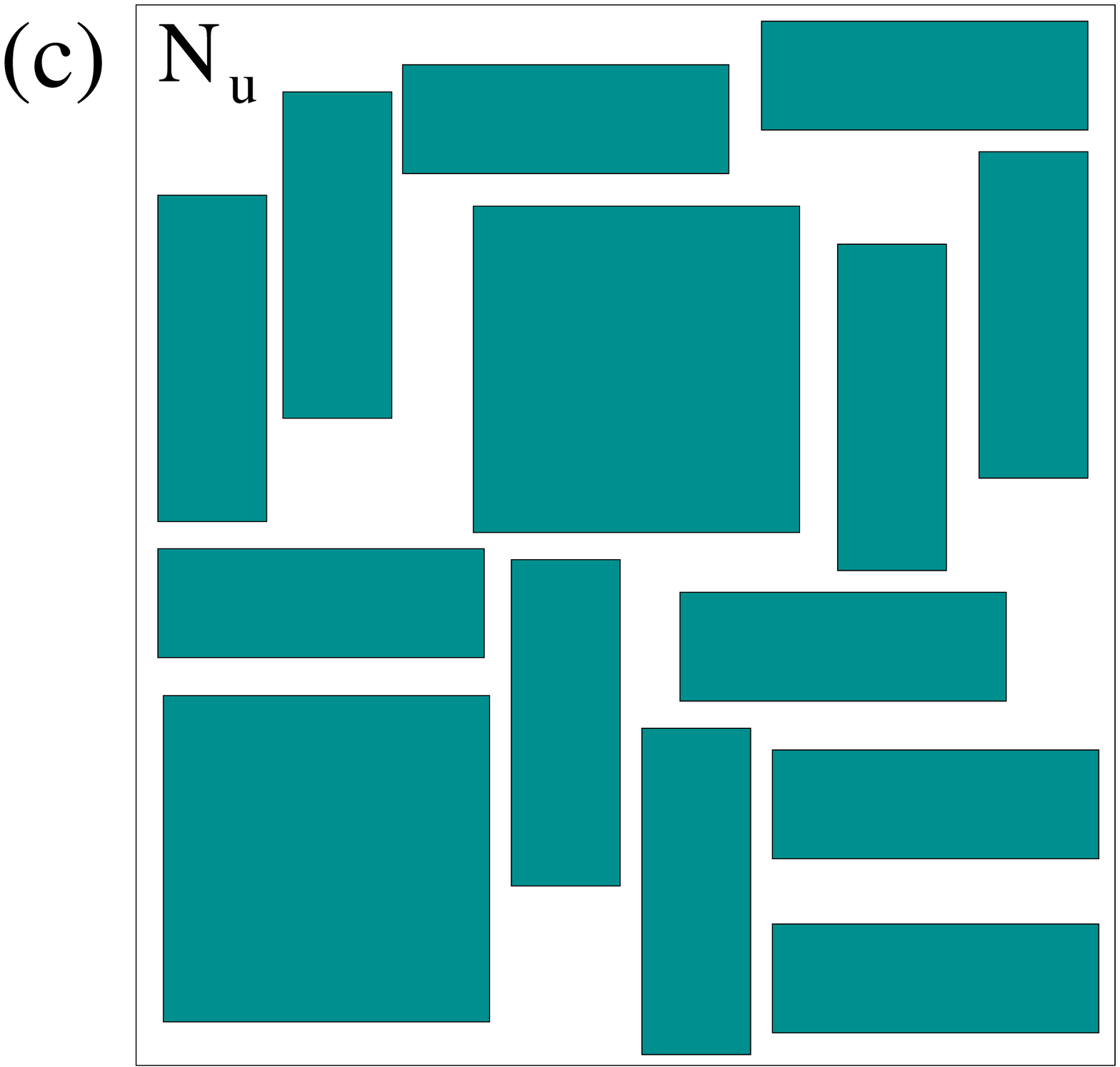,width=1.5in}
\hspace*{0.5cm}
\epsfig{file=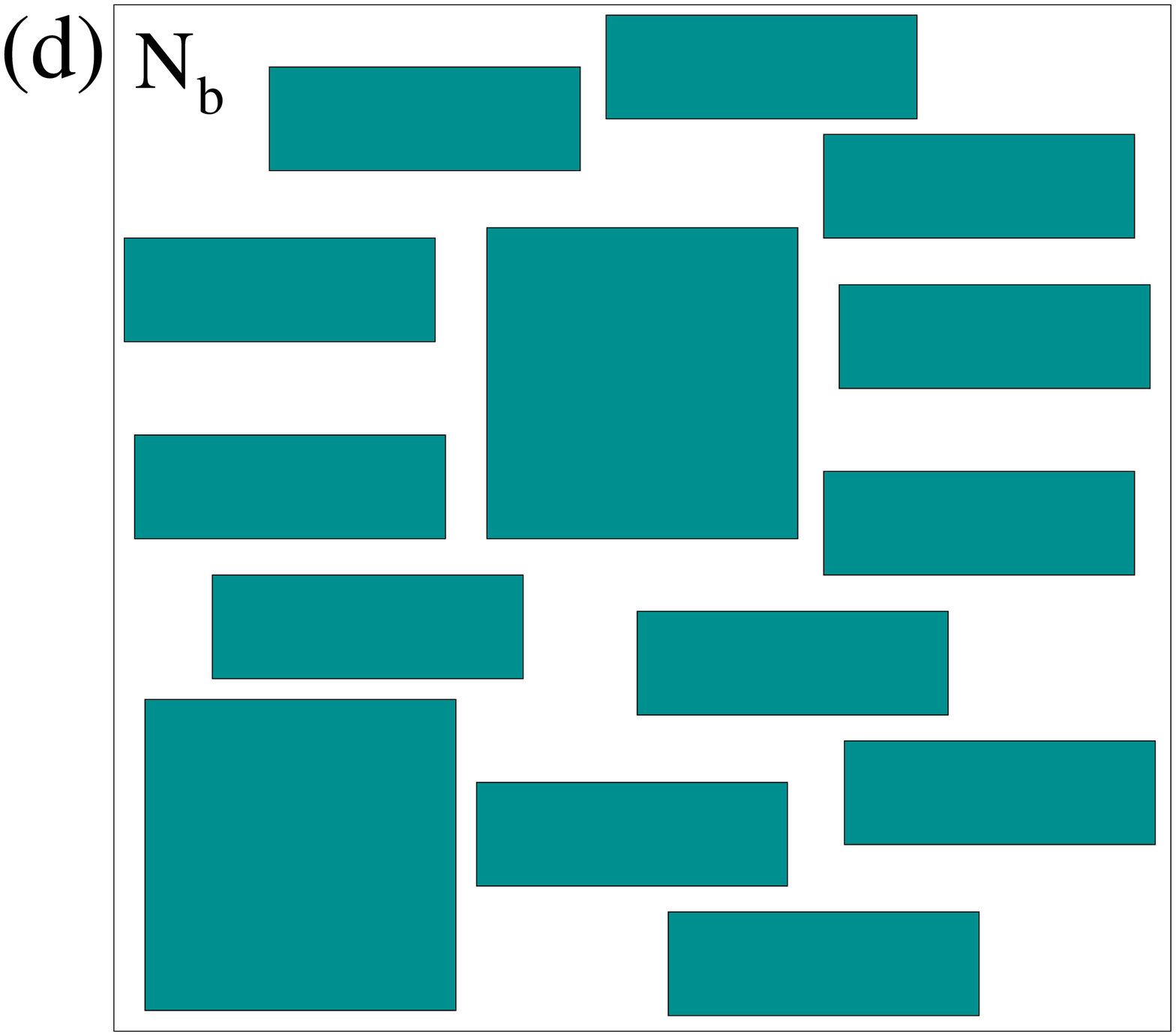,width=1.5in}
\caption{Sketch of cross-sectional particle configurations in the N$_{\rm u}$ phase [(a) for prolate and (c) for oblate particles] and 
N$_{\rm b}$ phase [(b) for prolate and (c) for oblate particles].}
\label{fig3}
\end{figure}

Fig. \ref{fig2}(a) contains the results obtained from the solutions of Eqns. (\ref{dos}) for the equilibrium order 
parameters $Q_{\rm u}$ and $Q_{\rm b}$ as a function of $\rho^*$ for particles with aspect ratios $1<\kappa\leq 20$. 
In this range of $\kappa$ we always find that
$Q_b=0$, i.e. the N$_{\rm u}$ is the only stable phase. As can be seen from the figure, the 
uniaxial order parameter $Q_{\rm u}$ increases from zero and saturates with $\rho^*$. The system does not exhibit
any orientational ordering phase transition (at zero density the order parameter departs from zero with a finite derivative). 
The order of rods builds up continuously with density along the direction perpendicular to the monolayer.
The N$_{\rm u}$ phase is depicted in Fig. \ref{fig3}(a), where the cross sections of 
particles are sketched: the most populated species corresponds to squares, while the 
probabilities to find rectangular species pointing along $x$ or $y$ are equal. 

The inset of Fig. \ref{fig2}(a) shows that $\rho^*$ is a monotonically decreasing function of $\kappa$ for fixed value
$0.3\le Q_{\rm u}\le 0.6$. As rods become longer, a fixed amount of nematic order requires a lower density. For higher 
values of $Q_{\rm u}$ this is not true (see main figure); however, when using $\rho_{\rm 2D}\sigma^2$ (mean particle number in 
a fraction $\sigma^2/A$ of the total area $A$) instead of $\rho^*$ as a 
density variable the trend is restored, since the ratio $\rho_{\rm 2D}\sigma^2/\rho^*=\kappa^{-1}$ decreases strongly
with $\kappa$. Finally the empty circle in Fig. \ref{fig2} represent the location of the N$_{\rm u}$-C,K 
bifurcation as calculated from the bifurcation analysis. We will return to this point later.  

In Fig. \ref{fig2}(b) we plot $Q_{\rm u}$ as a function of the packing fraction $\eta$. Now the curves do not intersect each other 
for any $\kappa$, but for large $\kappa$ ($\sim 20$) they exhibit a loop (see inset). The loop is not a signature of any phase 
transition, and is simply related with the dependence of $\eta$ with the order parameter $Q_{\rm u}$, as is explained in the 
following. From Eqn. (\ref{the_packing}) we can approximate $\displaystyle{\eta(Q_{\rm u})\approx\frac{2\rho^*}{3}(1-Q_{\rm u})}$ 
when $\kappa^{-1}\ll 1-Q_{\rm u}$. If $Q_{\rm u}\sim 1$, taking into account that 
$Q_{\rm u}$ is a monotonically increasing function of $\rho^*$, the 
lowering of $1-Q_{\rm u}$ can compensate, for certain values of $\rho^*$, the increment of the latter, giving a
decrease in $\eta$. Further, when $\kappa^{-1}\gg 1-Q_{\rm u}$, we have 
$\displaystyle{\eta(Q_{\rm u})\approx \frac{\rho^*}{3\kappa}(1+2Q_{\rm u})}$, an increasing function of $\rho^*$. 

\begin{figure}
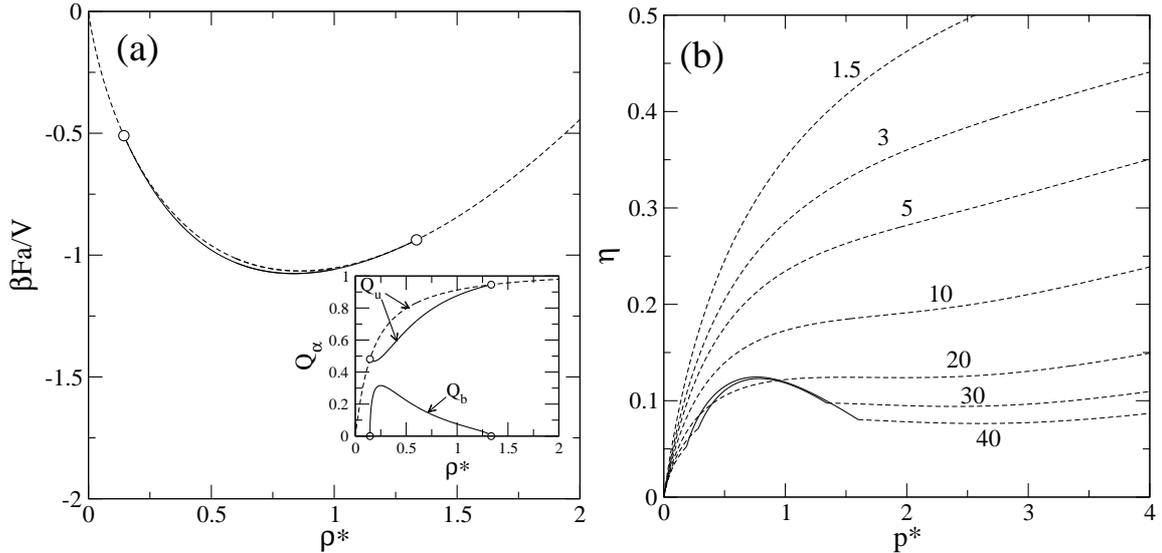

\epsfig{file=fig4a.eps,width=3.in}
\epsfig{file=fig4b.eps,width=2.9in}
\caption{(a) Free energies of the N$_{\rm u}$ (dashed curve) and N$_{\rm b}$ (solid curve) phases as a function of $\rho^*$ for $\kappa=40$. 
Inset: Order parameters Q$_{\rm u}$ and Q$_{\rm b}$ (correspondingly labelled) as a function of $\rho^*$ for the  N$_{\rm u}$ (dashed curve) and 
N$_{\rm b}$ (solid curve) phases. Open circles indicate the bifurcation points. (b) Packing fraction $\eta(Q_{\rm u})$ 
as a function of the reduced pressure $p*=\beta p a$ for different values of $\kappa$ (as labelled in the figure). 
The N$_u$ and N$_b$ branches are plotted with dashed and solid curves, respectively.} 
\label{fig4}
\end{figure}

When the aspect ratio $\kappa$ is high enough (for $\kappa>\kappa_0\simeq 21.339$) two solutions of 
Eqn. (\ref{bifurcation}) exist, associated with two N$_{\rm u}$-N$_{\rm b}$ bifurcation points. Therefore, the
sequence N$_{\rm u}$-N$_{\rm b}$-N$_{\rm u}$ is obtained as a function of density. To analyse this in more 
detail, the free energies of the N$_{\rm u}$ and N$_{\rm b}$ phases as a function of $\rho^*$ were calculated. The results 
are plotted in Fig. \ref{fig4}(a) for $\kappa=40$. Both  N$_{\rm u}$-N$_{\rm b}$ transitions are continuous.
The inset shows the order parameters $Q_{\rm u}$ and 
$Q_{\rm b}$ as a function of $\rho^*$. The mean-field behaviour $Q_{\rm b}\sim |\rho^*-\rho^*_{\rm b}|^{1/2}$ as
$\rho^*\to\rho^*_{\rm b}$ is confirmed, where $\rho^*_{\rm b}$ corresponds to any of the two bifurcation values.
In Fig. \ref{fig3}(b) the particle cross sections for the N$_{\rm b}$ phase are sketched in a situation where 
the majority of rectangular species point to the $x$ axis, $Q_{\rm b}\sim\gamma_x$ (i.e. the biaxial parameter is almost saturated; 
note that it can only saturate in a system of plates).

We can explain the presence of 
the reentrant $N_{\rm b}$ phase as follows. The strictly 2D fluid composed of hard-rectangular particles exhibits a 
continuous I-N$_{\rm u}$ transition at a packing fraction that decreases with aspect ratio as $\eta_{\rm 2D}=2/(\kappa+\kappa^{-1})$
(in the Zwanzig approximation). For the 3D Zwanzig rod fluid with centers of mass constrained on a plane, the packing fractions of 
parallelepipeds with rectangular projection, $\eta_{xy}\equiv\eta_x+\eta_y$, and square projection, $\eta_z$, can be computed as    
\begin{eqnarray}
\eta_{xy}(Q_{\rm u})\equiv \left(\rho_x+\rho_y\right)L\sigma=\frac{2\rho^*}{3}(1-Q_{\rm u}), \label{eta_xy}\quad 
\eta_z(Q_{\rm u})\equiv \rho_z \sigma^2=\frac{\rho^*}{3\kappa}(1+2Q_{\rm u}).
\end{eqnarray}
For highly elongated particles the total packing fraction can be approximated as 
$\eta(Q_{\rm u})\simeq \eta_{xy}(Q_{\rm u})$. In turn this quantity could coincide with $\eta_{\rm 2D}$ for some values of $\rho^*$  
(note that $Q_{\rm u}$ is a function of $\rho^*$ as obtained from the free-energy minimization with 
respect to the order parameters). When this occurs the  N$_{\rm u}$-N$_{\rm b}$ transition takes place. 
Note that, from (\ref{eta_xy}), the equality $\eta_{xy}=\eta_{\rm 2D}$ could be reached for two different values of $\rho^*$,
$\rho^{*(i)}$ ($i=1,2$) with $\rho^{*(1)}<\rho^{*(2)}$ and $Q_{\rm u}(\rho^{*(1)})<Q_{\rm u}(\rho^{*(2)})$ i.e., for  
two different fluids with different density and orientational order. 

We expect the 
same phase behavior for freely rotating rods. In particular a reentrant N$_{\rm b}$ phase might be found for high enough 
aspect ratios and in the range of packing fractions when the inequality $\rho A(Q_{\rm u})>\eta_{2D}$ is fulfilled. Here
$A(Q_{\rm u})$ is the mean particle area obtained by projecting the volume of a particle that forms an angle $\theta$ with respect 
to the layer normal on the layer perpendicular to the nematic director, and averaging with respect to the equilibrium angular 
distribution function $h(\theta)$ corresponding to the
equilibrium order parameter $Q_{\rm u}$. $A(Q_{\rm u})$ certainly depends on $\rho^*$, since $Q_{\rm u}$ is a function of $\rho^*$. 
In Fig. \ref{fig4} (b) we plot the packing fraction $\eta$ as a function of 
pressure for different values of $\kappa$ in the range $1.5\leq \kappa\leq 40$. For $\kappa<10$ the 
curves are always concave, while for higher values of $\kappa$ concavity is lost in some range of pressures 
(at high enough pressures concavity is recovered). This behavior is a direct consequence of the 
loop exhibited by the packing fraction $\eta(Q_{\rm u})$ as $\rho^*$ is increased, as 
explained above. For $\kappa>\kappa_0$ the curves 
exhibit a clear maximum in the region where the N$_{\rm b}$ phase is stable (see the solid curves corresponding to $\kappa=30,40$). 
When the rotational symmetry is broken in the $xy$ plane there is a clear gain in
surface area per particle, i.e. more particles can be packed with N$_{\rm b}$ symmetry.

\begin{figure}
\epsfig{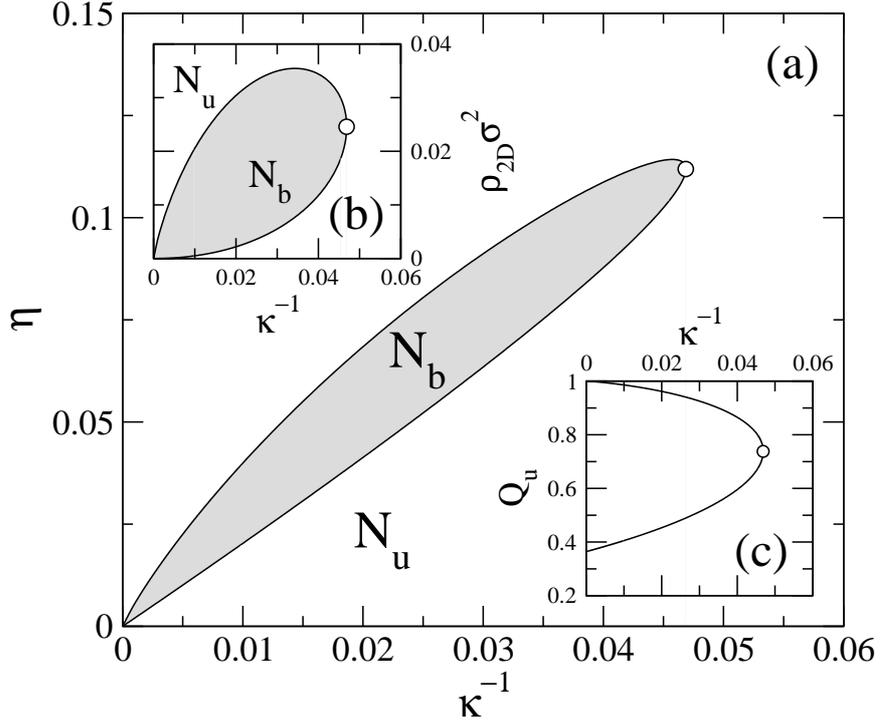}
\caption{(a) The N$_{\rm u}$-N$_{\rm b}$ bifurcation values of the packing fraction $\eta$ as a function 
of the inverse aspect ratio $\kappa^{-1}$ found from the solution 
of Eqn. (\ref{bifurcation}). (b) Bifurcation values for $\rho_{\rm 2D}\sigma^2$. (c) Uniaxial 
order parameter $Q_{\rm u}$ along the N$_{\rm u}$-N$_{\rm b}$ spinodal.} 
\label{fig5}
\end{figure}

The bifurcation values of packing fraction $\eta$ and scaled density $\rho_{\rm 2D} \sigma^2$,  
as obtained from the solution of (\ref{bifurcation}), are plotted in Fig. \ref{fig5}(a) and (b) as a function of $\kappa^{-1}$. 
As mentioned before, no solutions of Eqn. (\ref{bifurcation}) exist when $\kappa<\kappa_0$ since $\eta_z$, the 
packing fraction of hard squares, is of the order of $\eta_{xy}$ and the gain in excluded volume obtained 
by orienting the rectangular species along $x$ ($\gamma_x>\gamma_y$) is not enough to 
compensate the loss in mixing entropy. As can be seen from Fig. \ref{fig5}(b), the behavior of the lower and upper bifurcation curves 
are different in the Onsager limit $\kappa\to\infty$. Note that, for $\kappa\gg 1$, Eqn. (\ref{bifurcation}) reduces to 
\begin{eqnarray}
y\kappa^2-2=e^{2y\kappa+1}. 
\label{esta}
\end{eqnarray}
The lower branch can be obtained from (\ref{esta}) by defining the new variable $\theta=y\kappa^2$ and taking the limit 
$\kappa\to\infty$, which results in the solution $\theta=2+e$. Therefore the asymptotic behaviour of the lower branch is
$y_{\rm l}\approx \rho_{\rm 2D}^{(l)}\sigma^2=(2+e)/\kappa^2$. To find the asymptotic behavior of the upper branch we 
define another variable, $\tau=y\kappa$. Note that $\tau\to\infty$ when $\kappa\to\infty$, so that Eqn. (\ref{esta}) becomes 
$\tau\kappa=e^{2\tau}$. A fixed point algorithm 
$\displaystyle{\tau_n=\frac{1}{2}\ln(\kappa \tau_{\rm n-1})}$ with initial guess $\tau_0=1/2$ provides
\begin{eqnarray}
\tau_n=\frac{1}{2}\left\{\ln(\kappa/2)+\ln(\ln(\kappa/2))+{\cal O}\left[\frac{\ln(\ln(\kappa/2))}{\ln(\kappa/2)}\right]\right\},
\end{eqnarray}
which, for $\kappa\gg 1$, asymptotically gives 
\begin{eqnarray}
\tau^*=\lim_{n\to\infty} \tau_n\sim \frac{1}{2}\left\{\ln(\kappa/2)+\ln(\ln(\kappa/2))\right\}.
\end{eqnarray}
Thus we have $y_{\rm u}\approx\rho_{\rm 2D}^{(u)}\sigma^2=\tau^*/\kappa\to 0$ when $\kappa\to\infty$. It is interesting to note that 
$y_{\rm l}\sim (3+e)x^2$ and $\displaystyle{y_{\rm u}\sim -\frac{x}{2}\ln(2x)}$ 
(we have defined $x=\kappa^{-1}$). Then we obtain $y'_{\rm l}(x)\sim 2(2+e)x$ and $\displaystyle{y'_{\rm u}\sim -\frac{1}{2}\ln(2x)}$ 
for the first derivatives of these functions. Thus we find $y'_{\rm l}(x)\to 0$ and $y'_{\rm u}(x)\to\infty$ when $x\to 0$. 
These limits, together with the limits $y_{\rm l,u}(x)\to 0$, can be checked to hold numerically from Fig. (\ref{fig5})(b) (note that 
$\rho_{\rm 2D}\sigma^2\sim y$ for $\kappa\gg 1$). 
Taking into account Eqn. (\ref{packk}) we find 
that $\eta_{\rm l}(x)\sim 2x\to 0$ and $\displaystyle{\eta_{\rm u}(x)\sim -\frac{x}{2}\ln(2x)\to 0}$, when $x\to 0$, 
for the lower and upper branches of packing fractions at bifurcation.
In the same limit we have $\eta'_{\rm l}(x)\sim 2$ and $\displaystyle{\eta'_{\rm u}(x)\sim -\frac{1}{2}\ln(2x)\to\infty}$, which 
again can be verified from Fig. \ref{fig5}(a). In the inset (c) we plot the uniaxial order parameter Q$_{\rm u}$ along the 
N$_{\rm u}$-N$_{\rm b}$ bifurcation, as obtained from Eqns. (\ref{bifurcation}) and (\ref{ordering}). 
It is interesting to note the relatively low value of the order parameter ($Q_{\rm u}=0.36417$) at the point where the N$_{\rm b}$ solution 
first bifurcates in the Onsager limit ($\kappa\to\infty$).  

\begin{figure}
\epsfig{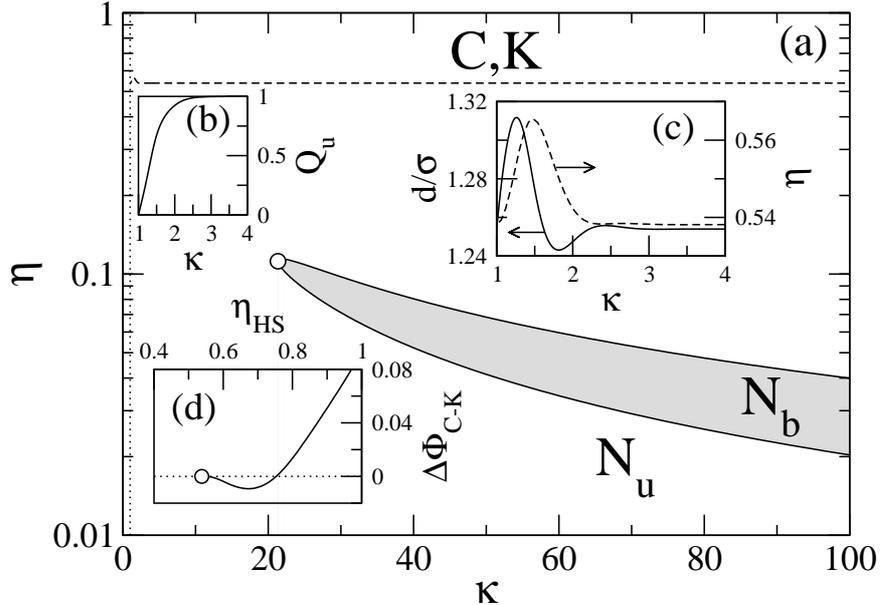}
\caption{(a) $\eta-\kappa$ phase diagram of rods in the Zwanzig approximation. The stability region of the N$_{\rm b}$ 
phase is shaded in grey and bounded by the solid line. The symbol 
label the critical end-point. The dashed line represents the N$_{\rm u}$-(C,K) spinodal. (b) Uniaxial order 
parameter $Q_{\rm u}$ along the latter spinodal.
(c) Detail of the N$_{\rm u}$-(C,K) packing fraction spinodal (dashed curve) and period in reduced units of the nonuniform phases at
bifurcation (solid curve).
(d) The free-energy difference between the C and K phases ($\Phi\equiv \beta {\cal F} a/V$, 
$\Delta \Phi_{\rm C-K}=\Phi_{\rm C}-\Phi_{\rm K}$) as a function of the two-dimensional packing 
fraction, $\eta_{\rm HS}=\rho_{\rm HS} \sigma^2$, for a one-component fluid of hard squares of sizes $\sigma$.} 
\label{fig7}
\end{figure}

We have also performed a bifurcation analysis to study the instability of the nematic against spatially nonuniform fluctuations,
such as columnar (C) or crystal (K) fluctuations. We solved Eqns. (\ref{dos}) and (\ref{structure}) to find the 
values of packing fraction $\eta$, order parameter $Q_{\rm u}$ and wave number $q=2\pi/d$ (with $d$ the periodicity of the density 
modulation along a given direction) at bifurcation. The results are plotted in Figs. \ref{fig7}(b) and (c) for $1\leq\kappa\leq 4$. 
As can be seen, the spinodal values of the packing fraction do not change 
too much as $\kappa$ is varied [see dashed line in panel (c)]. The same behaviour occurs with the period in reduced units $d/\sigma$ 
as a function of $\kappa$ [solid line in panel (c)]. This periodicity corresponds to that of the C or K phases with a 
high proportion of hard squares 
[see the evolution of the order parameter $Q_{\rm u}$ along the spinodal in panel (b)]. For larger values of $\kappa$, as $Q_{\rm u}\to 1$, 
the packing fraction tends asymptotically to that of the N-(C,K) bifurcation 
of the one-component fluid of hard squares, i.e. $\kappa=1$ (note that the C and K phases bifurcate at the same point). 

All of the above results, i.e. those for the N$_{\rm u}$-N$_{\rm b}$ continuous transition (Fig. \ref{fig5}), and those from 
the bifurcation analysis to nonuniform phases, are collected in Fig. \ref{fig7}(a), which is the complete phase diagram in the $\eta-\kappa$ plane. 
The only remaining question is the relative stability of the C and K phases at high densities. For the one-component fluid of hard squares, 
both phases bifurcate at $\eta_{\rm HS}\equiv \rho_{\rm HS}\sigma^2=0.5381$. The C is more up to $\eta_{\rm HS}\sim 0.75$,
beyond which the K phase becomes stable up close packing [see Fig. \ref{fig7}(c)]. In our system (a ternary mixture of hard squares 
and mutually orthogonal hard rectangles) and for high enough $\kappa$ (in particular for $\kappa >4$), we expect the same phase 
behaviour due to the small fraction of rectangular species at densities close to the spinodal instabilities (when $Q_{\rm u}\sim 1$). 
For $1\leq \kappa\leq 4$ a free-energy minimization of the ternary mixture in needed to elucidate 
this question. However we expect that the C phase will be the most stable phase, due to the fact that it is difficult to pack a ternary 
mixture of anisotropic species within a regular crystalline lattice.

\subsection{Oblate particles}
\label{oblate}

\begin{figure}
\epsfig{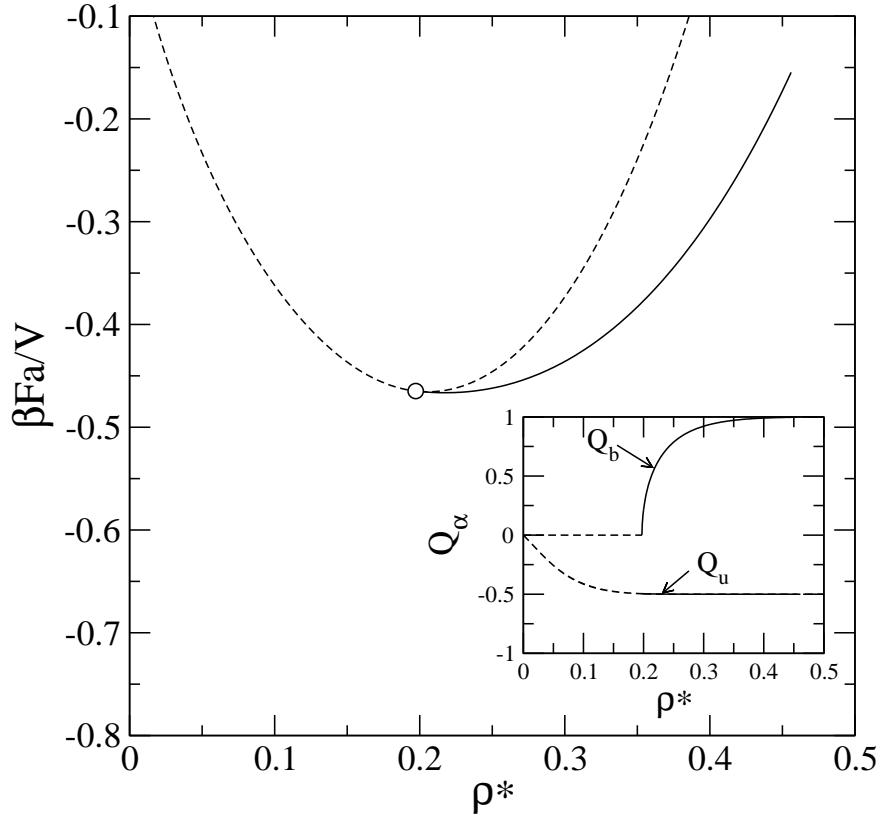}
\caption{Free energies of N$_{\rm u}$ (dashed curve) and N$_{\rm b}$ (solid curve) phases as a function of $\rho^*$ for $\kappa^{-1}=10$. 
The open circle indicates the bifurcation point. Inset: The evolution of the order parameters $Q_{\rm u}$ and $Q_{\rm b}$ 
as a function of $\rho^*$ along the N$_{\rm u}$ (dashed curve) and N$_{\rm b}$ (solid curve) branches.} 
\label{fig8}
\end{figure}

The phase behavior of oblate particles is characterized by a much wider region of stability of the N$_{\rm b}$ phase. This 
in turn can be explained by resorting to Figs. \ref{fig2}(c) and (d). Now the projection of the parallelepipedic species 
$z$ (now $\sigma>L$ and consequently $\kappa<1$) form squares with surface 
area $\sigma^2$. The other two species, $x$ and $y$, are rectangles with mutually orthogonal orientations of side-lengths 
$\sigma$ and $L$, with particle areas ($L\sigma$) much lower than that of the $z$-species, in particular when $\kappa$ is small enough. 
Therefore, as density is increased, the averaged excluded volume is lowered when the main particle axis lies in the $xy$ 
plane, and the uniaxial order parameter $Q_{\rm u}$ decreases from zero and tends asymptotically to $-0.5$ for high enough 
densities. Furthermore, since the system is quasi-two-dimensional, it exhibits a N$_{\rm u}$-N$_{\rm b}$ continuous phase transition 
at packing fractions $\sim \eta_{\rm 2D}$ (the packing fraction of the strictly two-dimensional fluid of hard rectangles). 

\begin{figure}
\epsfig{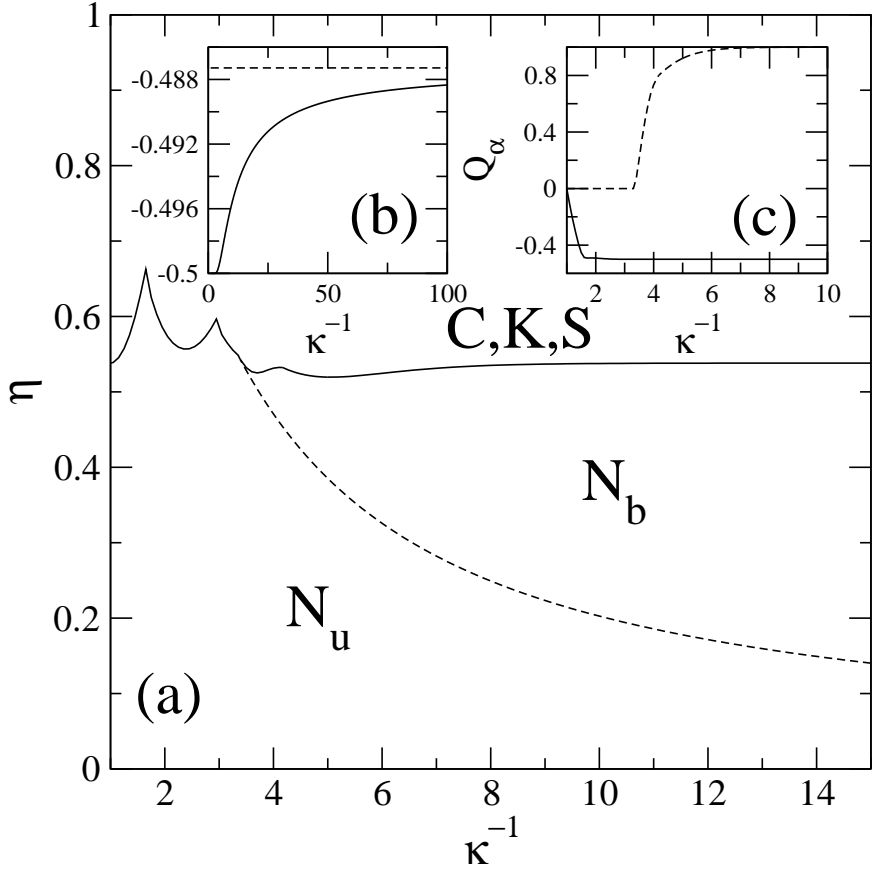}
\caption{(a) $\eta-\kappa$ phase diagram of rods in the Zwanzig approximation. The dashed line represents the location of the 
continuous N$_{\rm u}$-N$_{\rm b}$ transition, while the spinodal curve for the N$_{\rm u,b}$ instability against non-uniform (C, K or S) phases 
is shown with solid line. (b) Uniaxial order parameter $Q_{\rm u}$ along the N$_{\rm u}$-N$_{\rm b}$ transition (solid curve) and its 
asymptotic value for $\kappa\to 0$ (dashed curve). (c) Uniaxial (solid curve) and biaxial (dashed curve) order parameters along the spinodal 
curve corresponding to nonuniform phase instabilities.} 
\label{fig9}
\end{figure}

This behavior is shown in Fig. \ref{fig8}, where the free-energy branches of the $N_{\rm u}$ and $N_{\rm b}$ phases are 
plotted as a function of $\rho^*$. The inset shows the evolution of the order parameters $Q_{\rm u}$ and $Q_{\rm b}$ as a 
function of the same variable. $Q_{\rm b}$ departs from zero at the $N_{\rm u}$-N$_{\rm b}$ bifurcation point. Beyond this 
point the N$_{\rm b}$ is the most stable phase up to densities corresponding to the instabilities of the N$_{\rm b}$ phase against
spatially nonuniform fluctuations. These densities are calculated via the bifurcation analysis of Sec. \ref{bifurc}, and are plotted in the phase 
diagram of Fig. \ref{fig9}(a). Note the strong oscillations exhibited by the spinodal curve when $1\leq \kappa^{-1}\leq 4$ [i.e. the 
interval in which the biaxial order parameter $Q_{\rm b}$ is zero, see panel (c)]. 
This is because of the vanishingly small fraction of squares, with the rectangular species being
parallel to the $x$ or $y$ axes with equal probability. Thus, when the particle anisotropy grows, it becomes more difficult 
to commensurate the characteristic 
lengths of both rectangular species within the single lattice parameter of a nonuniform phase [C, K or smectic (S)]. The oscillations 
could have different origins: (i) when the ratio between the long and short edges of the rectangles is 
close to an integer number, square clusters can be formed by joining the rectangles along
their long sides. These clusters in turn can be accommodated into an square crystalline superlattice. (ii) The transitions to 
nonuniform phases could change from continuous to first order, with the coexisting density corresponding to the N$_{\rm u}$ phase 
located well below that estimated from the bifurcation analysis. (iii) The relative stability of C, S and K phases could 
strongly change with $\kappa$. In Fig. \ref{fig9}(b) the uniaxial order parameter $Q_{\rm u}$ along the 
N$_{\rm u}$-N$_{\rm b}$ bifurcation is plotted. It is important to note that, in the asymptotic limit $\kappa\to 0$, we have 
$Q_{\rm u}\to -0.4873$, i.e. biaxial ordering appears in particle configurations with a residual proportion of 
hard squares.    

\begin{figure}
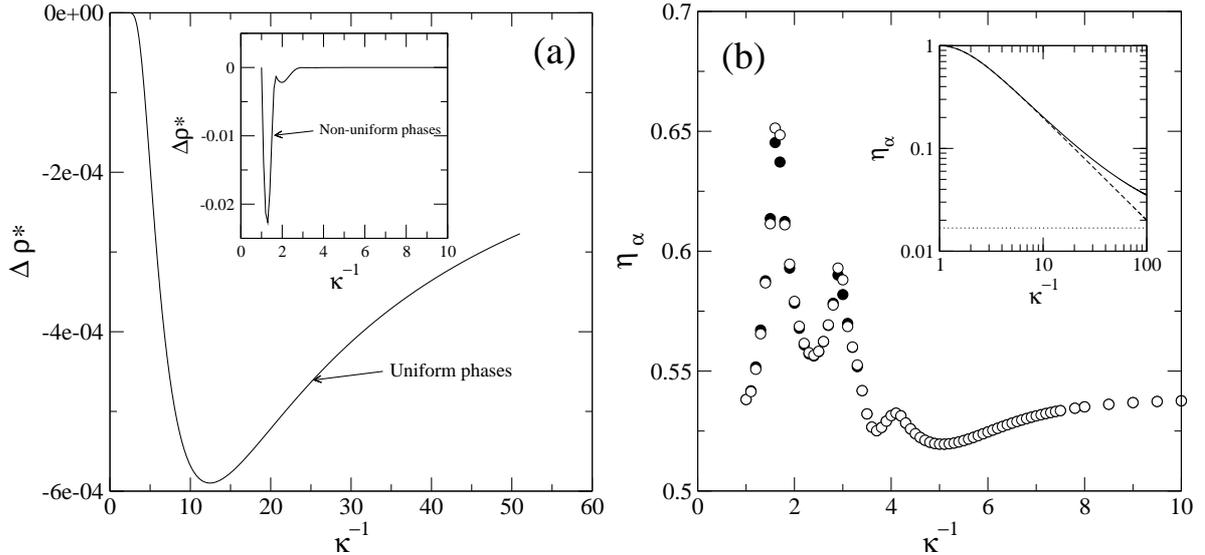

\epsfig{file=fig9a.eps,width=3.1in}
\epsfig{file=fig9b.eps,width=3.0in}
\caption{(a) Difference $\Delta \rho^*$ between the transition and spinodal curves of confined plates. 
Main figure: Difference between the density of the N$_{\rm u}$-N$_{\rm b}$ transition in 3D and that of 
the I-N$_{\rm u}$ transition in 2D. Inset: Difference between the N$_{\rm u,b}$-(C,K,S) spinodals in 3D and 
(I,N$_{\rm u}$)-(C,K,S) spinodals in 2D. (b) Packing fractions $\eta$ corresponding to the instability with respect to nonuniform phases for the 
monolayer of plate-like particles (empty circles) and for a strictly two-dimensional model (solid circles). Inset: The packing 
fractions corresponding to the N$_{\rm u}$-N$_{\rm b}$ (solid) and I-N$_{\rm u}$ (dashed) transitions for the monolayer and the two-dimensional 
model, respectively. The asymptotic packing fraction $\eta_a$ (see the text) is shown with dotted line.}
\label{fig10}
\end{figure}

Finally, in Fig. \ref{fig10}(a) we plot $\Delta \rho^*$, i.e. the difference between the spinodal 
curves corresponding to the three-dimensional Zwanzig oblate particles with their centers of mass located on a plane and 
the strictly two-dimensional system of Zwanzig hard rectangles. As we can see, the N$_{\rm u}$-N$_{\rm b}$ transition 
curve practically coincides with that of the I-N$_{\rm u}$ in 2D. The difference between the spinodal curves corresponding 
to transitions to nonuniform phases is larger for $1<\kappa^{-1}<2$, i.e., in the region where the strong 
oscillations in the spinodal take place. In the main figure of Fig. \ref{fig10}(b) we plot, for comparison,
the nonuniform phase spinodals corresponding to the plates monolayer (empty circles) 
and to the two-dimensional hard rectangles (solid circles). The 
conclusion is that they are practically the same. The inset in panel (b) shows the N$_{\rm u}$-N$_{\rm b}$ spinodals for both systems. 
The main difference is related to the asymptotic limit $\kappa\to 0$. For the two-dimensional system, the usual Onsager 
result $\eta\to 0$ results, while for the monolayer of plates the system retains a residual packing fraction $\eta_a=0.01681$, 
a result directly connected with the non-perfect uniaxial ordering ($Q_{\rm u}\neq -0.5$) of plates, as already discussed above. 

\section{Conclusions}
\label{conclusions}

Taking advantage of the $3D\to 2D$ dimensional crossover property of the fundamental measure density functional for the 
Zwanzig model, we have studied the phase behaviour of a monolayer of rod-like or plate-like uniaxial particles. We have focused, in particular, 
on their orientational ordering properties. Despite the fact that particles are uniaxial (they have only two characteristic lengths $L$ and 
$\sigma$), we have shown the presence of a N$_{\rm u}$-N$_{\rm b}$  phase transitions
for both particle geometries. For rod-like geometry, the N$_{\rm b}$ is a reentrant phase, since its region of stability in the phase diagram  
(aspect ratio-density plane) is always surrounded by that of the N$_{\rm u}$ phase. For this geometry the  
N$_{\rm u}$-N$_{\rm b}$ transition is continuous and the two N$_{\rm u}$-N$_{\rm b}$ transitions bounding the region 
of N$_{\rm b}$ stability meet at the critical end-point $(\kappa_0,\rho^*_0)$. For $\kappa<\kappa_0$ only the N$_{\rm u}$ 
is stable. For particles with large aspect ratio, there is inversion in the packing fraction with respect to the uniaxial order
parameter. For plate-like geometry, there exists a single continuous transition line separating the regions of N$_{\rm u}$ and 
N$_{\rm b}$ phase stabilities. This line crosses the spinodal instability to nonuniform phases at higher densities. 
The crossing point is approximately located at $\kappa\approx 0.3$. Therefore, below this aspect ratio, the N$_{\rm b}$ is stable at 
high enough densities until a point where stability with respect to C, S or K phases is lost. Above this aspect ratio the N$_{\rm u}$ is 
the only stable orientationaly-ordered phase;  at higher densities it again looses stability with respect to nonuniform phases. 
We have shown that the phase diagram of monolayers of plate-like particles is very similar to that of the strictly two-dimensional 
hard rectangular fluid in the restricted-orientation approximation.    

At present MC simulations of freely-rotating hard oblate ellipsoids with centers of mass confined on a plane are being carried out 
\cite{Gerardo}. We expect to find a qualitative agreement between the results obtained from the present model and the simulations.  

\acknowledgments
We gratefully acknowledge illuminating discussions with G. Odriozola.   
Financial support from Comunidad Aut\'onoma de Madrid (Spain) under the R$\&$D Programme of Activities 
MODELICO-CM/S2009ESP-1691, and from MINECO (Spain) under grants FIS2010-22047-C01 and FIS2010-22047-C04 is acknowledged.
SV acknowledges the financial support of the Hungarian State and the European Union under the TAMOP-4.2.2.A-11/1/KONV-2012-0071.

\end{document}